\documentclass{optica-article}

\journal{opticajournal} % for journals or Optica Open

\articletype{Research Article}

\usepackage{lineno}
\usepackage{soul}
\usepackage{graphicx} % Required for inserting images

   \usepackage[english]{babel}   % "babel.sty" + "french.sty"

  \usepackage{times}			% ajout times le 30 mai 2003
  \usepackage[font=small,labelfont=bf]{caption}
\usepackage{graphicx}
 \graphicspath{ }
\usepackage{mathtools}
\usepackage{epsfig}
\usepackage{graphicx}
 \usepackage{booktabs}
 \usepackage{multirow}
\usepackage{pifont}
\usepackage{caption}
\usepackage{subcaption}

\usepackage{array}
\usepackage{color}
\usepackage{colortbl}

\usepackage{pifont}
\usepackage{latexsym}

\usepackage{booktabs}
\usepackage{graphicx}
 \graphicspath{{/images/} }
\usepackage{mathtools}
\usepackage{epsfig}
\usepackage{graphicx}
 \usepackage{booktabs}
 \usepackage{multirow}
\usepackage{pifont}
\usepackage{caption}
\usepackage{subcaption}

\usepackage{xr}
\makeatletter

\newcommand*{\addFileDependency}[1]{% argument=file name and extension
\typeout{(#1)}% nlatexmk will find this if $recorder=0
\@addtofilelist{#1}
\IfFileExists{#1}{}{\typeout{No file #1.}}
}\makeatother

\newcommand*{\myexternaldocument}[1]{%
\externaldocument{#1}%
\addFileDependency{#1.tex}%
\addFileDependency{#1.aux}%
}
\myexternaldocument{Supplementary}
\begin{document}

\title{High-power femtosecond molecular broadening and the
effects of ro-vibrational coupling}

%Repetition rate tunable HCF pulse compression in N2 around 200W average power

\author{Kevin Watson,\authormark{1} Tobias Saule\authormark{1}, Maksym Ivanov\authormark{2,3}, Bruno E. Schmidt\authormark{2}, Zhanna Rodnova\authormark{1}, George Gibson\authormark{1}, Nora Berrah\authormark{1}, and Carlos Trallero\authormark{1,*}}

\address{\authormark{1}Physics Department, University of Connecticut, Storrs, CT 06269, USA\\
\authormark{2}few-cycle Inc., 1650 Blvd Lionel-Boulet, Varennes, Quebec J3X~1P7, Canada\\
\authormark{3}Institut National de la Recherche Scientifique (INRS), 1650 Blvd Lionel-Boulet, Varennes, Quebec J3X~1P7, Canada}

\email{\authormark{*}Corresponding author: carlos.trallero@uconn.edu} %% email address is required; see note below about the corresponding author designation

% use {asbstract*} to suppress the copyright line. Copyright information will be added in production

\section{Abstract} 

%Composit abstract 
%Scaling spectral broadening to higher pulse energies and average powers, respectively, is a critical step in ultrafast science, especially for narrowband Yb-based solid state lasers which become the new state-of-the-art. Despite their high nonlinearity, molecular gases as the broadening medium inside hollow-core fibers have been limited to 25 W, at best.

%Brunos Abstract
Scaling spectral broadening to higher pulse energies and average powers, respectively, is a critical step in ultrafast science, especially for narrowband Yb-based solid state lasers which become the new state-of-the-art. Despite their high nonlinearity, molecular gases as the broadening medium inside hollow-core fibers have been limited to 25 W, at best. We demonstrate spectral broadening in nitrogen at ten-fold average powers up to 250W with repetition rates from 25 to 200kHz. The observed ten-fold spectral broadening is stronger compared to the more expensive krypton gas and enables pulse compression from 1.3ps to 120fs. We identified an intuitive explanation for the observed average power scaling based on the density of molecular ro–vibrational states of Raman active molecules. To verify this ansatz, spectral broadening limitations in O2 and N2O are experimentally measured and agree well. On these grounds we propose a new perspective on the role, suitability, and limits of stimulated Raman scattering at high average and peak powers. Finally, high harmonic generation is demonstrated at 200~kHz.

%Origional abstract 
%Scaling spectral broadening to higher pulse energies and average power is a critical step for many laser systems including most Yb-based lasers which have become the new standard in tabletop ultrafast optics. Molecules have so far been unused for spectral broadening above 25~W due to Raman induced heating. In this paper $N_2$ is used for spectral broadening at a record 250~W of average power, providing an order of magnitude pulse compression to 120~fs with repetition rates of 25-200~kHz. Additionally, the transmission and broadening behavior of molecular gases $N_2$, $O_2$, and $N_2O$ is studied with repetition rates ranging from 25~kHz to 1~MHz and average powers from 100 to 250~W. We propose a new perspective on the suitability and limits of stimulated Raman scattering for pulse broadening at high average and peak powers. High harmonic generation is demonstrated at 200~kHz as a verification of pulse quality.

\section{Introduction}
%Control of the state of matter through light has been one of the ultimate goals in laser science. Perhaps one of the most impressive applications of light-matter interaction is the control of superconductivity, going as far as creating a path for near room temperature superconductivity. \cite{budden2021evidence} On the other hand, the largest phase transition recorded to date is the metallization of fused silica by femtosecond pulses. Both have in common the need for high intensity laser pulses. 
Since the discovery of chirped pulse amplification\cite{strickland1985compression}, Ti:Sapphire technology has been the workhorse of ultrafast optics, producing high peak powers, alas with limited average powers. %For some applications high average power ultrafast sources are advantageous due to limitations to one event per pulse, such methods include photoelectron spectroscopy \cite{gibson1989observation,mikaelsson2020high} and attosecond pump probe \cite{gibson1989observation, Attoseond_HHG_aplications_li2020attosecond}. 
Most if not all optical applications will benefit from higher duty cycles, which translates to higher repetition rates. For strong field science, this means achieving both high peak and average powers. 
%As well as a plethora of high-power ultrafast applications including advanced material processing\cite{fs_mech_for_QM_corrielli2021femtosecond,Aplications_MM2_lei2020ultrafast, aplicationsofUltrafastLasers_MM_sugioka2014ultrafast, Micro_mecheinging_samad2012ultrashort} and remote atmospheric sensing.\cite{Sensing_with_Filimentation_qi2022sensing, Filiment_sencor_inSpace_dicaire2016spaceborne, filiment_LIDAR_in_AIR_yu2001backward, 12km_Filiemntation_rairoux2000remote, ShortPulse_AIR_la1999filamentation} 
Recently the development and wide adoption of Ytterbium (Yb) based lasers over Ti:Sapphire has made the 100~W ultrafast regime accessible\cite{YbvsTi_Article_henrich2020ultrafast}. 
%This additional power enables much higher repetition rates, increasing the speed of data collection while lowering photoelectron space charge effects.\cite{saule2019highChargeEffects,Furch_2022} 
%As well as a plethora of high-power ultrafast applications including advanced material processing\cite{fs_mech_for_QM_corrielli2021femtosecond,Aplications_MM2_lei2020ultrafast, aplicationsofUltrafastLasers_MM_sugioka2014ultrafast, Micro_mecheinging_samad2012ultrashort} and remote atmospheric sensing.\cite{Sensing_with_Filimentation_qi2022sensing, Filiment_sencor_inSpace_dicaire2016spaceborne, filiment_LIDAR_in_AIR_yu2001backward, 12km_Filiemntation_rairoux2000remote, ShortPulse_AIR_la1999filamentation} 
However, Yb's narrow gain bandwidth produces long 1~ps-250~fs pulses \cite{YBGainBandwidth_honninger1999}. Therefore, to be used in ultrafast science these pulses must be broadened and compressed to drive many of the desired high intensity applications of ultrafast optics\cite{mcpherson1987studies,uiberacker2007attosecond,calegari2016advances,weissenbilder2022optimize,Attoseond_HHG_aplications_li2020attosecond,fs_mech_for_QM_corrielli2021femtosecond,Aplications_MM2_lei2020ultrafast, aplicationsofUltrafastLasers_MM_sugioka2014ultrafast, Micro_mecheinging_samad2012ultrashort,Sensing_with_Filimentation_qi2022sensing, Filiment_sencor_inSpace_dicaire2016spaceborne, filiment_LIDAR_in_AIR_yu2001backward, 12km_Filiemntation_rairoux2000remote, ShortPulse_AIR_la1999filamentation}. The most common methods of spectral broadening at high peak and average power are  noble gas filled multi-pass cells (MPC) or hollow core fibers (HCF) using self-phase modulation (SPM) \cite{nisoli2024hollow, multipassCellNeedForHighPowerMGStudy}.    % Added this review on HCF: DOI: 10.1109/JSTQE.2024.3373174 as suggested by Maks

%$\delta\omega(t)=-\frac{\delta \phi_{NL}}{\delta t}=-n_2\frac{L P_0 \omega_0}{c A_{eff}}\frac{\delta}{\delta t}|U|^2$

There is a growing interest in molecular gases as a broadening medium since many molecules exhibit a strong nonlinear refractive index $n_2$ \cite{ Molecular_Gas_Camparison_Milchberg2012} responsible for SPM induced spectral broadening: $\delta\omega(t)=-\frac{\delta \phi_{NL}}{\delta t} = -n_2 \frac{P_0 \omega_0 L_{eff}}{c A_{eff}}  \frac{\delta I}{\delta t}$ \cite{agrawal2000nonlinear}. Additionally, two effects from molecular structure have demonstrated multi-octave generation in HCFs and MPCs\cite{SRSselftrappingNat_safaei2020high, Makskumar2023generating, SuperConFromROT_beetar2020multioctave, Multipass_Cell, SuperContinMPS}. One effect is stimulated Raman scattering (SRS), in which photons interact with ro-vibrational states. Since molecular transitions are predominant in SRS an asymmetric red shifting is observed in the broadening \cite{ExtreamRScarpeggiani2020extreme,RamanINAIRli2014spectral}. SRS has historically been used to extend the frequency tunability of laser systems\cite{SRS_extendtunabilty_DUARTE2003157} and is still an active area of research with results such as SRS driven self-trapping of solitary states\cite{SRSselftrappingNat_safaei2020high}. The second effect is molecular alignment. As an ensemble of molecules interact with moderately intense pulses their polarizability axis aligns to the electric field's polarization, dramatically enhancing non-linearity\cite{SuperConFromROT_beetar2020multioctave, Fan:20,rodnova2023molecular}. The goal of high-power molecular broadening has until now focused on Raman red shifting . However previous experiments strongly suggest that molecular gases are not suitable for high-power applications as rotational heating suppresses broadening and transmission\cite{Molecular_Gas_Camparison_Milchberg2012,Thermal_Beetar:21}. Techniques to mitigate this heating have been implemented such as differential pressure schemes \cite{Dif_Pumpingarias2022few,SRSselftrappingNat_safaei2020high, Thermal_Beetar:21, Difrential} which have allowed functional molecular broadening up to 25~W.\cite{Dif_Pumpingarias2022few}

In this paper a molecular nitrogen ($N_2$) filled HCF broadening and compression scheme is implemented. For the first time, intput average powers up to 250~W were realized with repetition rates ranging from 25-200~kHz (corresponding to 12-1.5~mJ per pulse).  Pulse compression from 1.3~ps to 120~fs was achieved in this manner. This denotes an order of magnitude average power scaling compared to previous reports in molecular gas filled HCFs or MPCs. $N_2$ demonstrates 80~\% fiber transmission, spectral broadening similar and even slightly greater than Krypton ($Kr$), and good pulse compression while preserving high spatial quality with $M^2~=~1.2$. To understand this previously thought unreachable regime, the effect of high average power on molecular transmission and broadening is tested in $N_2$, molecular Oxygen ($O_2$), and Nitrous oxide ($N_2O$) at up to 250~W. An explanation of transmission loss from bandwidth driven cascaded SRS heating is developed based on these experimental findings and supported via molecular state calculations. Limitations of broadening in molecular gases in our experiments are shown to be a function of primarily pulse energy and result in a pulse breakdown. 
Pulse quality after broadening in $N_2$ with 218~W, 200~kHz is demonstrated by high harmonic generation in $Ar$. 

\section{Results}
%\section{high-power Pulse Broadening in Nitrogen}

Experiments were conducted with an \textit{Amphos} 3000 Yb amplifier system which delivers a maximum 300 W of average power. The pulses are broadened in a 750~$\mu$m diameter, 4.45~m long gas filled HCF (\textit{few-cycle Inc}). The versatile setup accepts input energies varying from 2mJ (200 kHz) up to 12mJ (25 kHz) at constant average power. Active beam pointing (\textit{TEM-Messtechnik}) is used to guarantee stable coupling into the fiber. The fiber output is compressed by a series of chipped mirrors (\textit{few-cycle Inc}). Full details of the experiment are shown in Fig.S1.  
 
First a study of pulse quality after $N_2$ broadening was conducted by optimizing gas pressure in the HCF for maximum pulse compression in the 185-250~W regime across 25-200~kHz. Almost identical SHG-FROG traces and power stability measurements of these compressed pulses can be found in Fig.S2 and S3. This HCF filled with $N_2$ displays high-power stability with 0.19~\% RMS power fluctuations over 2 hours of 3.92~mJ pulses at 196~W input power and 80~\% fiber transmission and continuously runs in the lab on a daily basis without degradation.

\begin{figure}[hbt!]
\centering\includegraphics[width=0.99\linewidth]{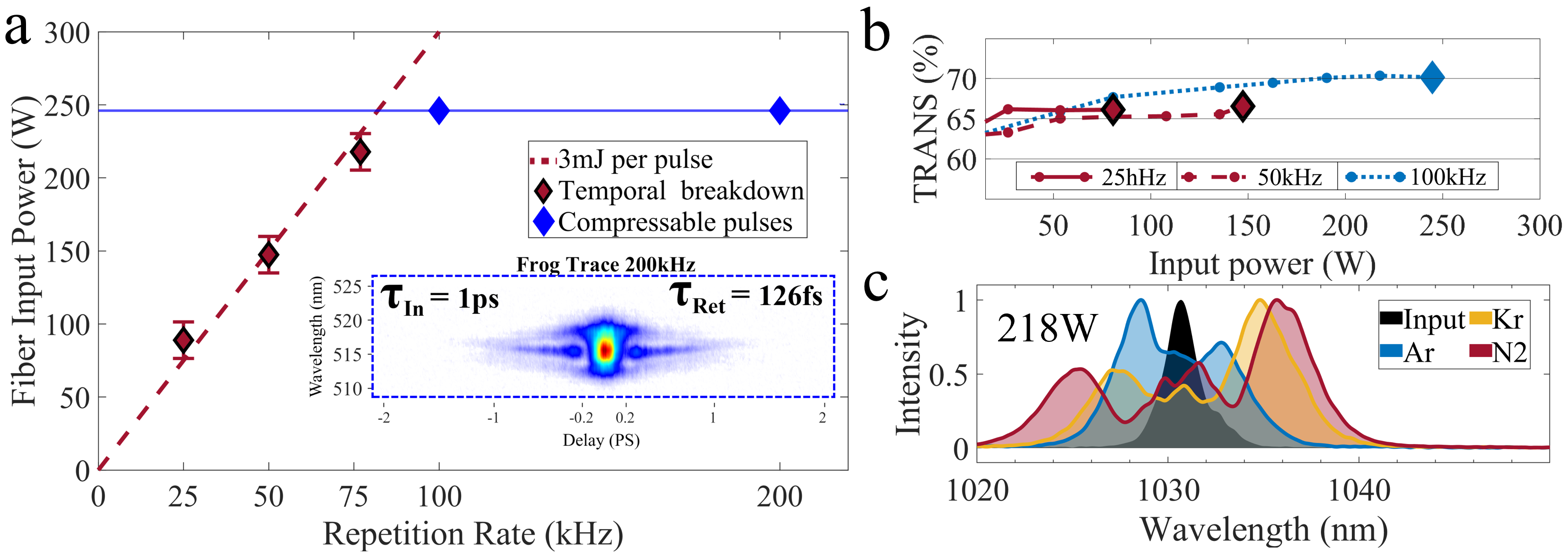} 
\caption{a) High average and peak power broadening in $N_2$ for different repetition rates and pulse energies demonstrating $\approx 120$~fs pulses at 218~W of input power with >70\% transmission  efficiency and a constant pressure of~1.5 bar. At repetition rates between 25-75~kHz temporal breakdown limits the maximum average power and occurs at 3~mJ of energy per pulse (red dotted line). Inset: SHG-FROG trace for 218~W, 200~kHz at 2.5~bar $N_2$ starting from 1.06~ps ($\tau_{In}$) to 126~fs ($\tau_{Ret}$) FWHM  of the intensity. b) HCF transmission at 25, 50, and 100~kHz repetition rates as a function of input power limited by temporal breakdown or maximum energy reached. c) Spectral broadening comparison of 218~W input power at 100~kHz in $N_2$, Argon ($Ar$) with 1.5~bar and Krypton ($Kr$) at 1.7~bar. }
\label{fig:N2_Fig}
\end{figure}

A second study of $N_2$ was conducted to understand the limitations of high-power molecular broadening. As the high-power regime is known to excite molecular rotational states resulting in transmission and bandwidth loss~\cite{Thermal_Beetar:21}, these qualities were measured in conjunction with pulse distortion for HCF broadening in $N_2$ with a constant pressure of 1.5~bar. Results are displayed in~Fig.\ref{fig:N2_Fig}. Output pulse distortion was measured though SHG-FROG measurements of temporal breakdown as described and displayed in Fig.~S4. Figure~\ref{fig:N2_Fig}a) shows the energy and power regimes where temporal breakdown was found to occur. Temporal breakdown was found to occur once 3~mJ per pulse was reached with no dependence on average power (dashed red line in panel a). For this study the gas pressure is kept constant at all repetition rates/powers. Therefore, it is surprising that $N_2$ behaves as a noble gas, whose pulse quality is impervious to repetition rate. Constant transmission of $\approx 70$~\% was sustained before the onset of temporal distortion of the compressed pulse at all repetition rates, as seen in Fig.{\ref{fig:N2_Fig}b)}. The bandwidth achieved in $N_2$ is shown in {\ref{fig:N2_Fig}c)} and it is $1.3\times$ wider at 1/e than that of $Kr$ at 1.7~bar. However, $Kr$ can operate at higher fiber pressure, near 2~bar for this case, before temporal breakdown of the output pulse. Simulations of nonlinear pulse propagation though $Kr$ and $N_2$ in a HCF using software Luna.jl produce bandwidths with strong agreement to our measurements as seen in Fig. S5 \cite{LUNA_brahms_2023_10323704}. Results for Ar are shown for reference. Temporal breakdown with increasing average power/energy per pulse was determined through SHG-FROG. Pulse degradation is detected in SHG-FROG scans as spectral and temporal instabilities. Temporal breakdown is typically the result from nonlinear self-focusing that distorts in synchronous the temporal and spatial properties of a pulse \cite{spatial_collapsecrego2019influence}.

%The bandwidth of $N_2$ grew continually with power to the spectra shown in Fig.\ref{fig:N2_Fig}c which is $1.3X$ wider at $\frac{1}{e}$ then $Kr$.

%These results are unexpected given previous research on high-power $N_2$ molecular pulse broadening. \cite{Thermal_Beetar:21}   

%Notice this $N_2$ spectra was driven by SPM not SRS as it is symmetric, without the red shifted center wavelength that results from Raman scattering. This observation is developed later to explain the seeming disagreement between our results and these previous experiments.  

%\section{Transmission in Other gasses} 
%\textcolor{red}{generalized sounds wrong here: maybe: these findings are in contrast to previous studies that were limited to 25W in $N_2$. To shed light onto the differences we conducted...}

\begin{figure}[hbt!]
\centering\includegraphics[width=\linewidth]{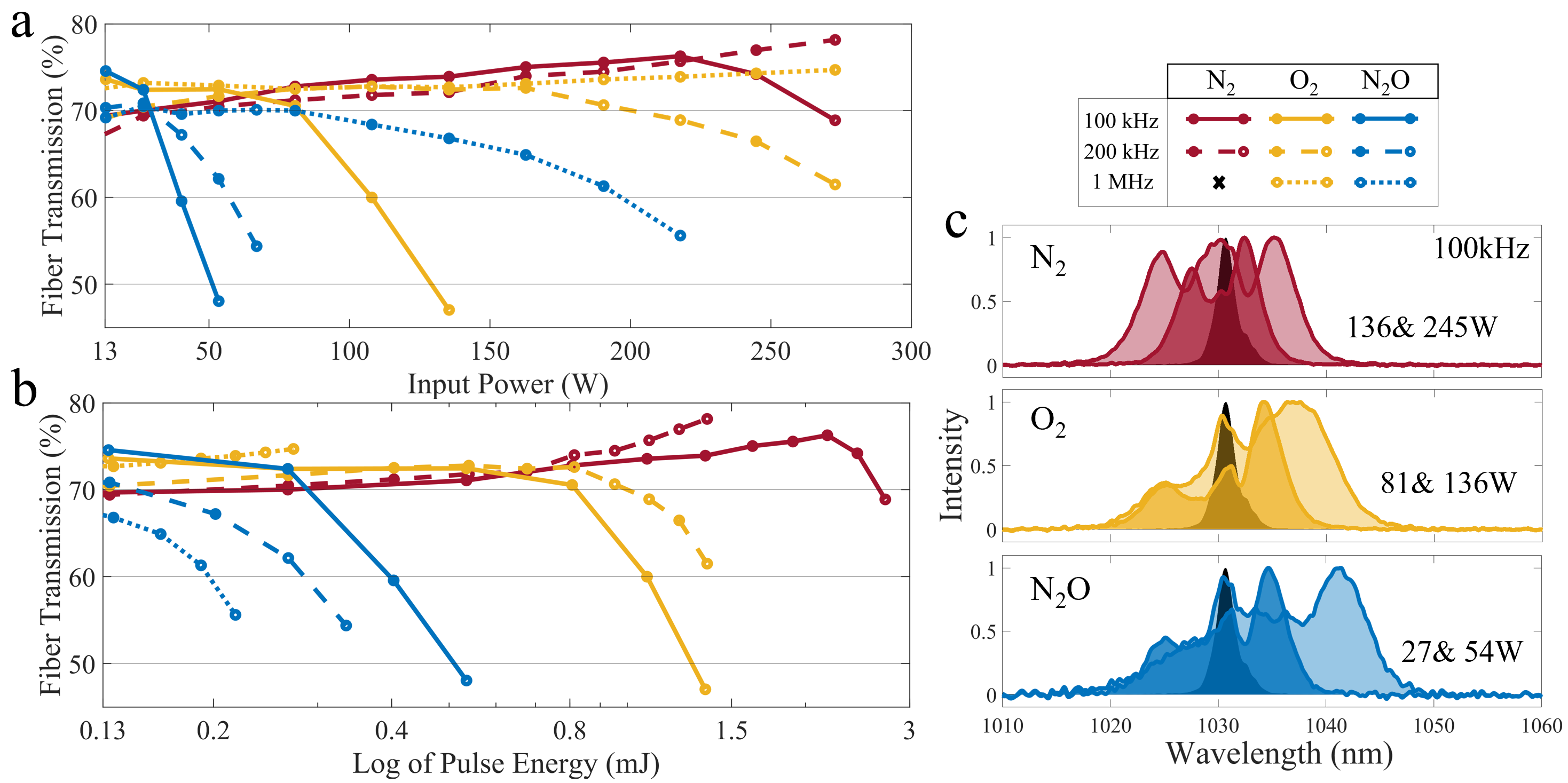} 
\caption{Comparison transmission efficiency and pulse broadening in gasses $N_2$, $O_2$, and $N_2O$  for 100~kHz, 200~kHz, and 1~MHz repetition rates. a) Transmission as a function of input power. b) Transmission vs log of pulse energy. Legend is on the top right. c) Symmetric spectra from $N_2$ at 136~W and 245~W. Largest symmetric and asymmetric spectra generated in $O_2$ at 81~W, 136~W and in $N_2O$ at 27~W, 136~W respectively. Black spectra are that of the input pulse.}
\label{fig:difrent_gasses}
\end{figure}

Our findings are in contrast to previous studies that generated broad SRS red shifted spectra but could not overcome average input power beyond 25~W ($N_2O$) for any molecular gas. To understand our results and their relation to other works we conducted a further transmission and broadening study of $N_2$ with two other molecular gasses, $O_2$ and $N_2O$, as illustrated by Fig.\ref{fig:difrent_gasses}. $O_2$ and $N_2O$ are useful comparisons, having similar ionization potentials but very distinct nonlinear indices and ro-vibrational state densities\cite{Molecular_Gas_Camparison_Milchberg2012, IonizationMolecules_tong2002theory}.

 At these high average powers for a given repetition rate $O_2$ undergoes transmission loss and mode breakdown sooner then $N_2$. Figure~\ref{fig:difrent_gasses}b) is the same data as in panel a), but plotted in energy per pulse rather than average power. As seen in Fig.\ref{fig:difrent_gasses}b), $O_2$ withstands over twice the pulse energy of $N_2O$ without undergoing transmission loss. This demonstrates strong field ionization is not the primary cause of transmission loss because $O_2$ has a slightly lower ionization energy at 12.07~eV than $N_2O$ with 12.89~eV. In log scale, it can clearly be seen that $N_2O$, $O_2$, and $N_2$ are capped in the amount of broadening at three very different ranges of energy per pulse. Figure \ref{fig:difrent_gasses}c) shows the broadened spectra compared to the input spectrum for all three molecules at two powers indicated in each panel. While the $N_2$ spectra remains symmetric, broadened spectra for $O_2$ and $N_2O$ are largely asymmetric, favoring red wavelengths. This is indicative of SRS-induced broadening, instead of SPM \cite{Extream_Raman_Redshift2020extreme}. $N_2$ is then free of significant SRS heating and therefor experiences transmission loss from pulse energy like a Nobel gas. On the other hand, molecules undergoing Stokes Raman transitions in ro-vibartional manifolds, lead to excitation. If such excitations are long-lived the molecule does not return back to the ground state in the time between pulses leading to residual heat being deposited. This repetition rate dependent heating explains the loss in transmission to pulse energy experienced by $N_2O$. A plausible explanation for loss in transmission is thermal lensing at the entrance of the fiber resulting in unreliable coupling  \cite{SRSselftrappingNat_safaei2020high}.

\begin{figure}[ht!]
\centering\includegraphics[width=13cm]{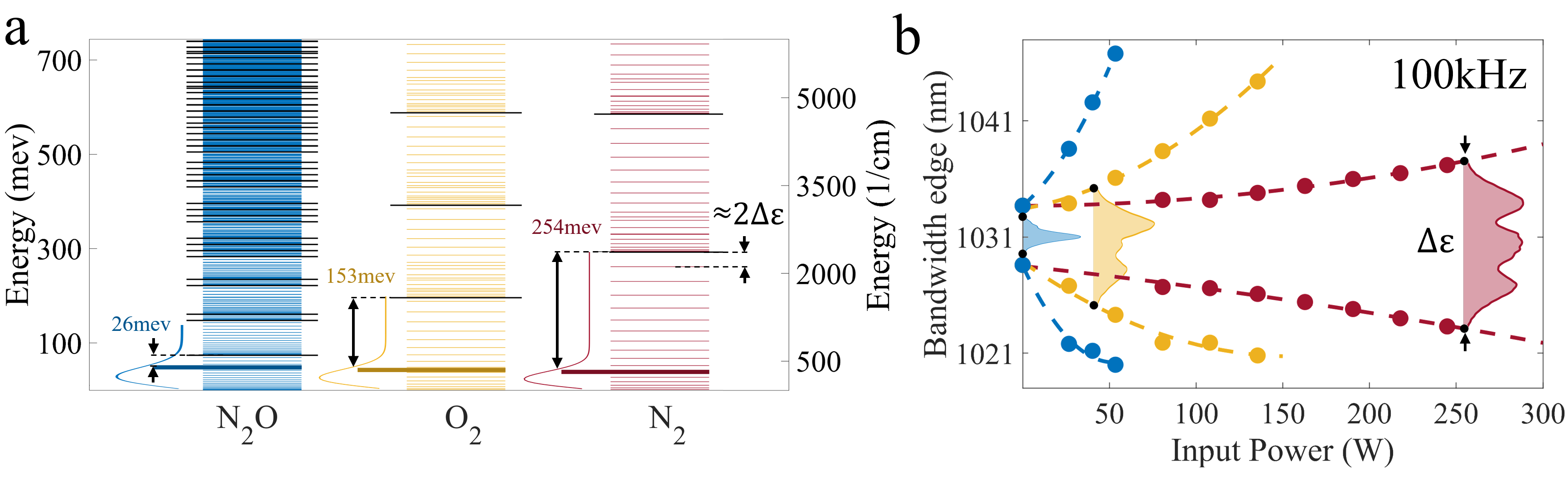} 
\caption{ a. Rotational-vibrational energy levels for molecular gasses $N_2O$, $O_2$, and $N_2$. First 20 odd number molecular rotational states  (Colored horizontal lines) super imposed on vibrational states (Wider Black lines). Only odd rotational states are plotted as Raman selection rules for linear molecules require $\Delta~J = \pm2$. The population distribution of each gas at room temperature (300~K) is displayed vertically on the left of every gas's density of states. Energy from 1/2 peak population at room temperature to first vibrational state is given. b. Bandwidth edges at $1/10$ max intensity of pulses broadened HCF with 1.5~bar of $N_2$ (red), $O_2$ (yellow), and $N_2O$ (blue) are shown as a function of power. Spectra corresponding to the largest ro-vibrational energy transition gap $\Delta \epsilon$ in each molecule is displayed at the power at $\Delta \epsilon$ bandwidth is reached.}
\label{fig:density of states}
\end{figure}

We propose an explanation of our results based on cascaded SRS molecular heating revealing previously unidentified high-power capabilities of molecules. The amount of SRS a molecule undergoes is directly influenced by the density of states in the vibrational manifold and the bandwidth of the laser. Stokes SRS processes couples two energy levels with two different photons from within the bandwidth of a single pulse. One photon is absorbed exciting a virtual state while the second photon drives stimulated emission. As seen in Fig.~\ref{fig:density of states}a) $N_2O$ has an exceptionally dense manifold of ro-vibrational states in comparison to $N_2$ and $O_2$ in addition to a substantially lower first vibrational level. An explanation of state density calculations is given in the supplementary \cite{herzberg1945molecular, johnson2013computational}. The dependence of SRS transitions on ro-vibrational state density and pulse bandwidth is shown in Fig.~\ref{fig:density of states} and reveals how the density of states clamps the maximum amount of bandwidth that can be generated before laser-induced ro-vibrational coupling occurs.

%Pargraph here 

%A molecules density of states is comprised of a manifold of superimposed rotational and vibrational energy levels, creating an state density that increases with every vibrational mode activation. 

We argue that molecular state density is relevant when the bandwidth is large enough to cover the spacing between the first excited vibrational level and the next lower lying rotational states as denoted by $\Delta \epsilon$ in Fig.~\ref{fig:density of states}. For gasses $N_2O$, $O_2$, and $N_2$ this clamping ro-vibrational transition $\Delta \epsilon =$ 3.8~meV, 11.8~meV, and 16.6~meV respectively, following rotational Raman transitions for linear molecules  $\Delta J=\pm 2$ \cite{RamanSpecRulesschrotter2001raman}. Therefore, $\Delta \epsilon$ acts as an intuitive benchmark for the minimum bandwidth needed to drive cascaded SRS. Once this climbing process starts, and rotational excitation reaches the first excited vibrational state internal conversion couples rotational states to the vibrational manifold\cite{bixon1968intramolecular,herman1996optical,kaufman2020coherent}. This process can repeat to reach higher vibrational states. For higher pulse repetition, the molecule remains in the vibrational states before it can relax and heat ensues. Therefor, as pulse bandwidth exceeds $\Delta \epsilon$ cascaded SRS heating becomes unbounded, capable of exciting vibrational states.

To further make this point, the progression of spectral broadening in each gas as a function of input power is displayed in Fig.~\ref{fig:density of states}b) and corresponds to the transmission data of Fig.~\ref{fig:difrent_gasses}. This allows a qualitative measurement of the prevalence of SRS in each gas at a given power through the degree to which a spectrum is red shifted. SPM broadening is symmetric for a pulse with temporal symmetry. On the other hand, self steepening deforms a pulse as the center peak moves slower though most media \cite{anderson1983nonlinear,SelfSteepeingdemartini1967self} yielding blue shifted broadening. We thus can determine the relative prevalence of self steepening, SPM, and SRS in a spectra by considering the direction of shifts in center wavelength. It is also of note that molecular alignment occurs due to Raman transitions between rotational levels, but the degree of alignment is diminished for vibrational excited states. Further, it has been shown that raising the temperature of all linear molecules increases the threshold intensity needed for alignment \cite{szekely2022alignment,kumarappan2006role}.

Figure~\ref{fig:density of states}b) clearly shows the predominance of red shifting in spectral broadening and how it differs greatly for each gas. To illuminate this, the $\Delta \epsilon$ of each molecule is given in bandwidth and displayed as a spectrum at the average power where the $\Delta \epsilon$ bandwidth is reached for each gas.  Red shifting is observed insignificant when pulse bandwidth is less then $\Delta \epsilon$. $N_2$ just reaches $\Delta \epsilon_{N_2}$ bandwidth at 250~W which we accredit to why $N_2$ spectra lacks noticeable SRS red shifting. While $N_2O$ has such a small $\Delta \epsilon$ that the input pulse is already broad enough to drive SRS transitions beyond the first excited vibrational level losses occur soon after. $O_2$ reaches $\Delta \epsilon_{O_2}$ of bandwidth near 50~W and is therefore able to accommodate greater powers than $N_2O$.  

Therefore, we submit that broadening in molecular-filled HCF occurs through both Kerr and SRS processes, but for bandwidth > $\Delta \epsilon$ ro-vibrational state climbing will take over and will clamp the amount of broadening attainable at high repetition rates. With this in mind, one can explain why we are able to couple hundreds of watts in average power compared to previous studies, since our starting pulses are of ps in duration vs 100-300 fs.

\begin{figure}[hbt!]
\centering\includegraphics[width=13cm]{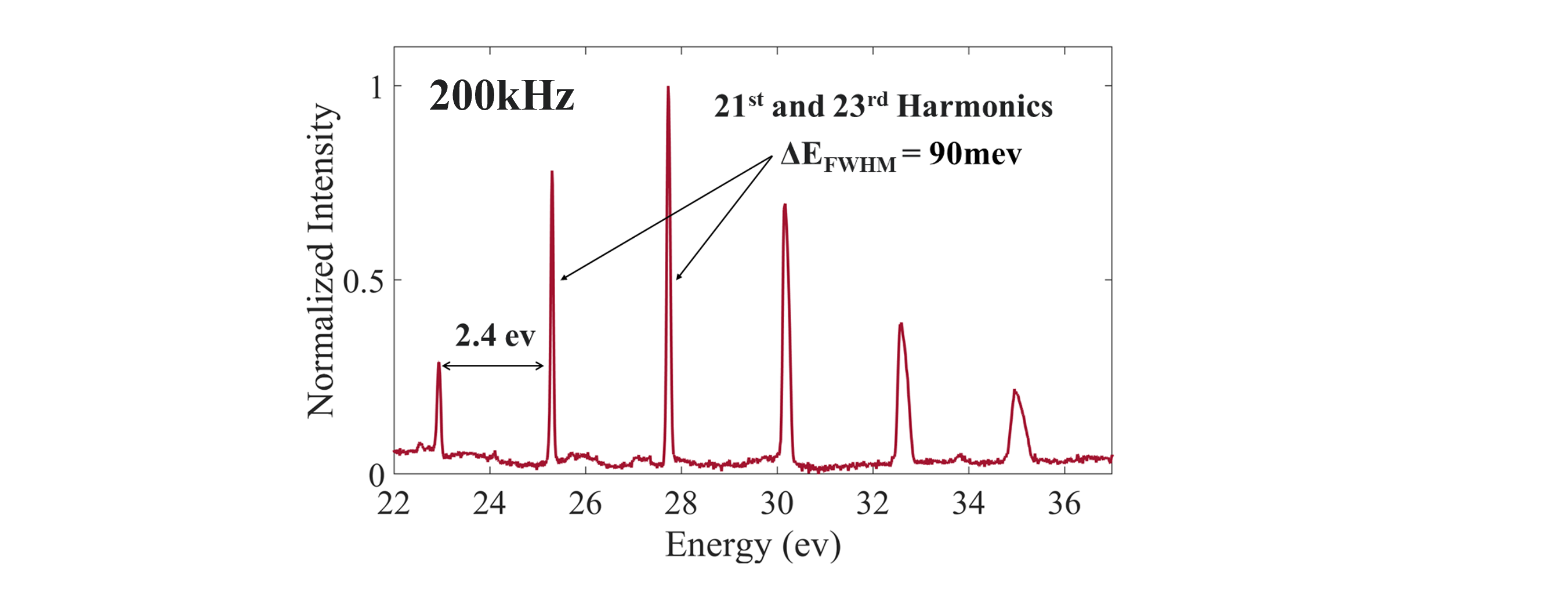} 
\caption{High harmonic generation spectra from 200kHz pulses shown in the FROG trace of Fig.\ref{fig:N2_Fig}a)}
\label{fig:HHG}
\end{figure}

One of the most direct demonstrations of pulse quality is through a very high non-linear process such as high-harmonic generation (HHG). In Fig.~\ref{fig:HHG} 52~W of a total of 160~W output power, broadened in $N_2$, and compressed to 136~fs at 200~kHz, were used to drive HHG in Ar. This leaves 108~W of additional compressed power for use in parallel experiments. As harmonics get sharper in energy with consecutive periodic sources our 35 optical cycle long pulses produce XUV pulses with bandwidth < 90~meV and we believe they are ideal for photoelectron spectroscopy experiments.

\section{Conclusion}

We have demonstrated pulse broadening and compression in a molecular-filled HCF at 10~times the previous recorded power for any molecular gas, reaching output powers of 218~W at 200~kHz. Input 1.3~ps pulses are compressed to 126~fs with 80\% hollow core fiber transmission with high pulse quality, demonstrated through HHG with only 52~W of power at 200~kHz. A comparison between the gases $N_2$, $O_2$, and $N_2O$ is conducted. This study yielded a framework for the interplay between bandwidth and ro-vibrational coupling as the limiting factor for broadening in molecular gases. More precisely, in the absence of SRS or when SRS is limited, molecules behave as atoms in terms of broadening. That is, broadening is limited by the input energy per pulse. For SRS-dominated broadening, the level spacing between the rotational levels of the ground vibrational manifold and the first excited vibrational state seems to be the limit in bandwidth that can be reached at high repetition rates, due to heat deposition in the molecule and consequent loss in transmission. This fact explains why it is possible to couple and broaden >250W of average power in a molecular-filled HCF with ps-long pulses and still achieve a factor of 10 in compression and high transmission efficiency.

Our findings can have strong implications for intense, high repetition rate, ps pulsed laser propagation in the atmosphere where the dominant species are $N_2$ and $O_2$. Therefore, we believe our findings will guide future directed energy laser atmospheric propagation studies and design.

\section{Acknowledgements}
This work was partially funded by an ONR Ultra-short Pulse laser and atmospheric characterization DURIP and regular grants N00014-18-1-2872 and N00014-19-1-2339 . KW was partially funded by AFOSR grant FA9550-21-1-0387. TS was partially funded by US Department of Energy,
Office of Science, Chemical Sciences, Geosciences, \& Biosciences Division grant DE-SC0024508. ZR was partially funded by a Directed Energy Professional Society grant. Additional supported was warded by the College of Libearl Arts and Sciences of the University of Connectocut. 
%This broadening is on the level of $Kr$ with a 80\% fiber termination, high pulse quality, and good compressibility at 10 times the previous record for molecular gas broadening in a HFC or multi-pass cell. A broadening and transmission study of $N_2$, $O_2$, and $N_2O$ reveals the role of SRS has in molecular heating at high-power. Allowable SRS transitions in these molecular gasses ro-vibrational state densities varies greatly with there density of states and SRS transitions are suppressed if allowable energy transitions are greater then pulse bandwidth. This new perspective guides future use of molecular gasses at high-power based on molecular density of states and the bandwidth of the laser being implemented. Applications of high-power $N_2$ broadening are demonstrated with $Ar$ HHG at 200~kHz and strong field ionization of $Kr$ for VMI studies at 50~kHz.

%

\bibliography{Works-Cited-KBW_HCF.bib}

\end{document}